\documentclass[a4paper,12pt]{article}

\usepackage{amssymb,amsmath}
\usepackage[pdftex]{graphicx}
\usepackage{indentfirst}

\newcommand{\ZZ}{\mathbb Z}
\def\lgl{\langle\hspace{-0.2em}\langle}
\def\rgr{\rangle\hspace{-0.2em}\rangle}
\newcommand{\deff}{\,\stackrel{\rm def}{\equiv}\,}
\title{Effect of Activity and Inter-Cluster Correlations on Information-Theoretic Properties of Neural Networks}
\author{A.~Demichev\\
{\small\textit{Skobeltsyn Institute of Nuclear Physics, Lomonosov Moscow State University,}}\\
{\small\textit{1(2), Leninskie gory, GSP-1, Moscow, 119991, Russia}}}
\date{}

\begin{document}

\maketitle

\begin{abstract}
On the basis of solutions of the master equation for networks with a small number of neurons it is shown that the conditional entropy and integrated information of neural networks depend on their average activity and inter-cluster correlations.
\end{abstract}

\section{Introduction}
It is well known that the real neural networks are subject to internal and external random influences (cf., e.g., \cite{SK}, \cite{PTW}, \cite{FSW}). Therefore, stochastic models of biological neural networks in recent years has attracted a lot of attention \cite{OC}, \cite{BC}, \cite{Bre} (a recent survey of stochastic methods in neuroscience can be found in \cite{LL}). In this paper we consider stochastic neural networks on the basis of the master equation \cite{Cow}, \cite{OC}, \cite{BC}.

The aim of this work is to explore the interdependence of information-theoretic metrics (ITM) for stochastic neural networks and the more simple quantities such as the first distribution moments of the activity of individual neurons. The ITMs are important for a quantitative description of the cognitive abilities of neural networks and are based on the concept of information entropy. Examples of the ITMs include conditional entropy, integrated information, effective information, stochastic interaction, etc. (see, e.g., \cite{BT2008}, \cite{BS}, \cite{NB}, \cite{AW}). However, even numerical calculation of such metrics for sufficiently large networks is an extremely challenging and resource-intensive task since their calculation requires knowledge of the full joint probability for the states of all the neurons (the probability of a given configuration of the network). Moreover approximate methods of calculation of these metrics are currently poorly developed. It is therefore very important to investigate the possibility of using for description of the cognitive properties  along with the above mentioned ITMs more simple  characteristic properties for which there exist well-developed methods for the approximate and numerical calculations.

The specific objective of this work is to establish an interdependence between such a hard-to-calculate ITM as the integrated information \cite{BT2008}, or, more precisely, the integrated information averaged over the states of the network (average integrated information; AII) \cite{BT2008}, \cite{BS}, and the relatively simpler parameters, namely, the average network activity, average correlations between the activity of individual neurons and inter-cluster correlations. For simplicity, we will call them correlation metrics (CM) for neural networks. The advantage of the CMs is that they are expressed in terms of finite order moments of the activity of neurons and, therefore, in the case of large networks it is much easier to develop approximate methods for their calculation. Unfortunately, the determination of the exact matching for ITMs and CMs by analytical methods is very difficult (if solvable at all) problem. It is well known that the recovery of the full joint distribution function for the states of all the neurons (needed for calculation of the ITMs) from its moments is a very complicated problem that does not have comprehensive solutions despite many years of research in this area (see., e.g., \cite{MH} and references therein). But in our case we are facing even more complex variant of this problem, namely, establishing a relationship between quantities constructed from the probability distribution function for neural network configurations, and the averages over this distribution of products of the activity variables of certain groups of neurons (see more details in section~\ref{sec:SND}).

Therefore, in this paper we try to establish a numerical relationship of these metrics for a certain set of neural networks. We consider networks of rather small size, with the number of neurons $N = 8$. Even for such a relatively small networks calculation of such ITMs as the conditional entropy (CE) and the average integrated information (AII) requires the solution of a large number of equations (a system of 256 differential equations for 256 initial conditions). This can be done with the help of computer algebra systems; in particular, in this work we used \textit{Mathematica} \cite {Wol}. As a result we have established a dependence of the CE and AII on neural correlations for such networks. We hope that in the future these results can be generalized to certain classes of neural networks of large size. In addition, the results for networks of small size may be of self-contained interest, for example when using ITMs for developing neuromorphic robots (cf., e.g., \cite{MHC}).

The paper is organized as follows. The section~\ref{sec:SND} briefly describes the model of the neural network used in this work. In section~\ref{sec:ITCQ} the ITMs and CMs are specified, and also the corresponding notations are introduced. In section~\ref{sec:VTKV} a methodology for calculating the ITMs and CMs of evolving stochastic neural networks are described, and as well the results of the calculations are presented. Section~\ref{sec:Zak} contains a discussion and conclusion.

\section{The neural network model: stochastic neurodynamics \label{sec:SND}}
As mentioned in the Introduction, we will consider the stochastic neural networks in the framework of stochastic neurodynamics (SND) \cite{Cow}, \cite{OC}, \cite{BC}. In this approach neural networks are considered as non-equilibrium stochastic Markov systems. The basis of the description is the master equation. More specifically, in our paper we consider a model with two-state neurons \cite{Cow}, \cite{OC}. In this version of the SND, neurons at each node of the network can be in one of two states, either ``active'' or ``quiescent'' (passive). It is assumed that the active state corresponds to firing and the quiescent to resting of a neuron. Synaptic connections are determined by the incidence matrix of the network $W = \{w_{ij}\}$. An active neuron spontaneously goes into a quiescent state with a decay rate $\alpha$. The rate of transition into the active state is determined by the activation function $f(w_ {ij}v_j)$.

A configuration of the networks is described by the random variables $v_i\in \ZZ_2,\ i=1,\dots,N$ ($N$ is the number of neurons in the network). More precisely, we assume that $v_i$ can take two values: $v_i=\{0,1\},\ i=1,\dots,N$. For convenience we introduce the notation ${\vec v}=\{v_1,\dots,v_N\}$. On the configuration space the probabilities $P(v_1,\dots,v_N;t)\equiv P({\vec v};t)$ are defined. They are determined from the master equation
\begin{equation}
 \frac{\partial P({\vec v};t)}{\partial t}=\sum_{{\vec v}^{\,\prime}\neq{\vec v}}\Bigl(\Omega_{{\vec v}{\vec v}^{\,\prime}}P({\vec v}^{\,\prime};t)-\Omega_{{\vec v}^{\,\prime}{\vec v}}P({\vec v}^{\,\prime};t)\Bigr)\ ,                                                                                   \label{SND1}
\end{equation}
where $\Omega_{{\vec v}{\vec v}^{\,\prime}}$ is the probability of the system transition from the configuration ${\vec v}^{\,\prime}$ to the configuration ${\vec v}$. 

Hereafter in this paper we consider the linear activation function
\begin{equation}
f(w_{ij}v_j)=f\cdot w_{ij}v_j\ ,                                                               \label{SND2b}
\end{equation}
where $f$ is a constant. Notice that in the stochastic case the assumption of linearity does not seem so far away from the real biological neurons, as in the case of deterministic models, where saturation effects are obviously very important.

In the case of a linear activation function the master equation (\ref{SND1}) can be written after the time rescaling  $f\cdot t\rightarrow t$ in the form
\begin{equation}
 \frac{\partial P({\vec v};t)}{\partial t}=\sum_i \Bigl[\lambda(v_i+1)+ v_iw_{ij}v_j\Bigr]P({\vec v}_{i+};t)
-\Bigl[\lambda v_i+(v_i+1)w_{ij}v_j\Bigr]P({\vec v}_{i};t)\ .                                                                                   \label{SND2c}
\end{equation}
where $\lambda=\alpha/f$, ${\vec v}_{i\pm}$ are the special configurations matching the ${\vec v}$, except that $i$-th component is $v_i \pm 1$. The addition is understood in the sense of $\ZZ_2$, i.e. by $\mod\,2$.

Further investigation of neural networks in the framework of this approach is based on solving the system of equations (\ref{SND2c}). It is obvious that for large networks an exact solution is not achievable and it is necessary to develop some approximate methods, the latter being a very difficult task \cite{OC}, \cite{OC2}, \cite{BC}. We will use the exact solution of the system (\ref{SND2c}) for networks of relatively small size.

\section{Information-theoretic and correlation metrics\label{sec:ITCQ}}

\subsection{Average Integrated Information}
As the main ITM, we consider the average integrated information defined below in this section. The very integrated information $\phi[V;\vec v]$ is defined as follows \cite{BT2008}:
\begin{equation}
\phi[V;\vec v]\deff \varphi[V;\vec v;{\mathcal P}^{MIP}]\ ,            \label{bs2}
\end{equation}
where
\begin{equation}
\varphi[V;\vec v;{\mathcal P}]\deff H\Bigl[P_{\vec V_0|\vec V_1=\vec v}\Bigm|\Bigm|\prod_{k=1}^r P_{\vec M^k_0|\vec M^k_1=\vec m^k}\Bigr]\ ,            \label{bs1}
\end{equation}
is the effective information and $\mathcal P^{MIP}$ is the partition for which the normalized effective information has a minimal value. In the definitions (\ref{bs2}) and (\ref{bs1}) we have used the following notations: $V$ is the system of binary neurons; $\vec V$ is the vector of random variables corresponding to the configuration of neurons ; $\vec v$ is a particular realization of the system configuration (the set of configurations of neurons); $\vec m^k$ is the configuration of $k$-th subsystem when all the system $V$ has the configuration $\vec v$; $H[\cdot||\cdot]$ is the relative entropy (Kullback-Leibler divergence);  $P_{\vec V_0|\vec V_1=\vec v}$ is the conditional probability distribution for the network at $t=0$ given that at $t=1$ the system is in the configuration $\vec V_1=\vec v$.

We denote the solution of equation (\ref {SND2c}) with the initial condition $P_0(\vec v_0)$ (distribution at the initial time $t_0$) as follows: $P(\vec v,t;[P_0(\vec v_0)])$. Let the complete system be divided into two parts: $V=\{M,L\}$. Accordingly, the vectors of  system configurations take the form: $\vec v=\{\vec m,\vec l\}$. The probability of the subsystem to be in the configuration $\vec m$ has the form
\begin{equation}
 P^M(\vec m,t;[P_0(\vec m_0)])= \sum_{\vec l,\vec l_0}P(\vec m, \vec l,t;[P_0(\vec m_0,\vec l_0)])P^N_{unif}(\vec l_0)\ ,    \label{uess2b}
\end{equation}
where $P^L_{unif}$ is the uniform distribution for the subsystem $L$. Thus, the probability distribution for a subsystem is obtained by summing over the final configurations of the subsystem and additional averaging over the initial states of the supplementary subsystem by using of the uniform distribution \cite{BS}.

As is usually done, in the formulas of the type (\ref{bs1}) we consider only the case of $r = 2$, i.e. bipartitions, because this greatly simplifies calculations. The effective information averaged over the network configurations has the form \cite{BT2008}, \cite{BS}:
\begin{equation}
\widetilde{\varphi}[V;t;\mathcal{B}]=\sum_{k=1}^2H_c(M_0^k|M_t^k)-H_c(V_0|V_t)\ . \label{bs7}
\end{equation}
where $H_c(V_0|V_t),\ H_c(M_0^k|M_t^k)$ are the conditional entropies for the corresponding neural (sub)networks. Recall that the conditional entropy $H_c(X|Y)$ determines the amount of information that is needed to describe the possible realizations of the random variable $X$ if we know the value of another random variable $Y$:
\begin{equation} 
H_c(X|Y) = \sum_{x\in\mathcal X, y\in\mathcal Y}p(x,y)\log \frac {p(y)} {p(x,y)}. \label{it6}
\end{equation}
Here $p(x,y)$ is the joint probability distribution and $p(y)$ is the marginal distribution; $\mathcal X$ and $\mathcal Y$ are the sample spaces for $X$ and $Y$, respectively. Further details can be found, for example, in \cite{CJ}.

The average integrated information (AII) is defined as follows
\begin{equation}
\widetilde\Phi[V;t]\deff\widetilde{\varphi}[V;t;\mathcal{B}^{MIB}]\ ,            \label{bs5}
\end{equation} 
where $\mathcal{B}^{MIB}$ is the minimal bipartition:
\begin{eqnarray}
\mathcal{B}^{MIB} &\deff& \mathrm{arg}_\mathcal{B}\min\left\lbrace\frac{\widetilde{\varphi}[V;t;\mathcal{B}]}{K(\mathcal{B})}\right\rbrace\ , \label{bs8} \\[3mm]
K(\mathcal{B})&\deff&\min\Bigl[H(M^1_t),H(M^2_t)\Bigl]\ .                   \label{bs9}
\end{eqnarray}
Here $H(M^{1,2}_t)$ are the subsystem entropies at the time $t$.

Further details on the definition and properties of AII can be found in \cite{BS} and \cite{BT2008}. There are also other information-theoretic metrics that may play a role similar to the AII, see, e.g., \cite{BS}, \cite{AW}, \cite{NB}.

\subsection{Correlation metrics}
Let us introduce the following notations for the mean values
\begin{itemize}
\item $\lgl\cdot\rgr$ denotes averaging over the stochastic process, i.e. averaging with the distribution function $P(\vec v,t;[P_0(\vec v_0)])$, which is the solution of the master equation (\ref{SND2c});
\item a symbol with a bar denotes a quantity averaged over the system (for example, $\bar{v}$ is the average activity of neurons in the network).
\end{itemize}

In this paper, we use the average of the variables of neural activity $v_i$ of order not higher than the second, i.e.,$\lgl v_i\rgr$ and $\lgl v_iv_j\rgr$ ($i,j=1,\dots,N$). The first one determines the average activity of the $i$-th neuron while the second determines correlation between the activities of the $i$-th and $j$-th neurons. From these averages one can form the well-known combinations:
\begin{itemize}
\item covariance
\begin{equation}
c_{ij}\deff \lgl v_iv_j\rgr - \lgl v_i\rgr\lgl v_j\rgr \ ;          \label{km}
\end{equation}
an important advantage of the covariance is that it vanishes for independent random variables;
\item since in general a magnitude of the covariance depends on measurement units, the Pearson correlation coefficient (PCC) $r_{ij}$ is widely used:
\begin{equation}
 r_{ij}\deff\frac{\lgl v_iv_j\rgr-\lgl v_i\rgr\lgl v_j\rgr}{\sigma(v_i)\sigma(v_j)}=\frac{c_{ij}}{\sigma(v_i)\sigma_t(v_j)}\ , \label{ksii2}
\end{equation}
where
\begin{equation}
\sigma(v_i)=\sqrt{\lgl v^2_i\rgr-\lgl v_i\rgr^2}  \label{ksii3a} 
\end{equation}
is the standard deviations. However for the binary variables $v_i$ in our case the dependence of the covariance on measurement units is irrelevant.
\end{itemize}

The corresponding quantities averaged over the system are defined as follows
\begin{itemize}
\item  average activity
\begin{equation}
\bar{v} \deff \sum_{i=1}^{N}\lgl v_i\rgr\ , \label{skh1}
\end{equation}
\item average second moment for the neuron variables
\begin{equation}
\bar{m}\deff \frac{2}{N(N-1)}\sum_{i>j}\lgl v_iv_j\rgr\ , \label{skh1a}
\end{equation}
\item average covariance
\begin{equation}
\bar{c} \deff \frac{2}{N(N-1)}\sum_{i>j}c_{ij}\ ,\label{skh2}
\end{equation}
\item average PCC
\begin{equation}
\bar{r}\deff \frac{2}{N(N-1)}\sum_{i>j}r_{ij}\ ,\label{skh3}
\end{equation}
\end{itemize}

In addition to the above mentioned conventional correlation metrics we will use the inter-cluster correlation coefficient (ICC):
\begin{eqnarray}
k_{S\hat{S}}&\deff& \frac{2 \bar{c}_{S\hat{S}}}{\left(\bar{c}_{S}+\bar{c}_{\hat{S}}\right)}    \label{WC1b}\\[3mm]
\bar{c}_{S}&\deff&\frac{1}{N_S(N_S-1)}\sum_{i,j\in S_K}c_{ij} \label{WC2a}\\[3mm]
\bar{c}_{S\hat{S}}&\deff&\frac{1}{N_SN_{\hat{S}}}\sum_{i\in S;j\in \hat{S}}c_{ij} \ ,
\end{eqnarray}
where $\hat{S}$ is the supplementary subnetwork for $S$: $S\subset V,\ \hat{S}=V\setminus S$, $V$ is the set of node of the network under consideration, $N_S,\ N_{\hat{S}}$ is the number of nodes in the corresponding subnetworks.

\section{Calculations ITMs and CMs for the stochastic neural networks and establishing interrelations between them\label{sec:VTKV}}

\begin{figure}
\begin{center}
\includegraphics[scale=.43]{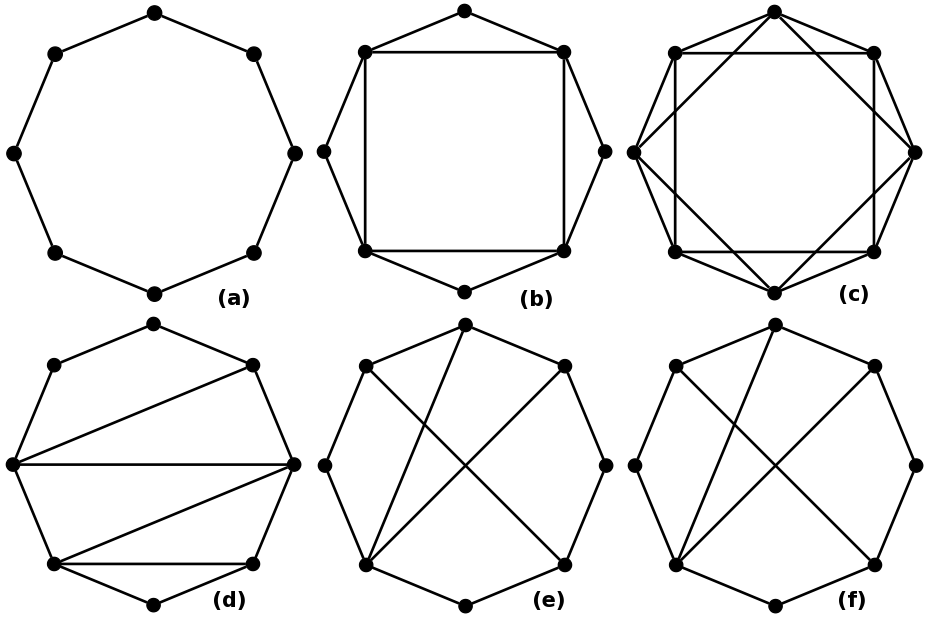}
\end{center}
\caption{The undirected networks with eight neurons, $N=8$, for which ITMs and CMs are compared}
\label{fig:combined_undirected}
\end{figure}

\begin{figure}
\begin{center}
\includegraphics[scale=.43]{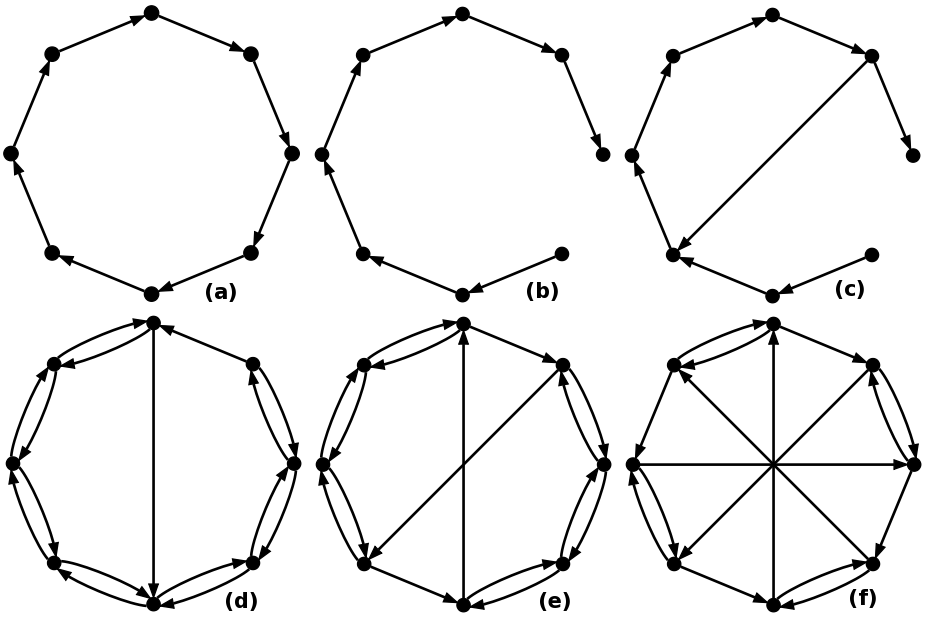}
\end{center}
\caption{The directed networks with eight neurons, $N=8$, for which ITMs and CMs are compared}
\label{fig:combined_directed}
\end{figure}

We consider a set of six undirected networks (Fig.~\ref{fig:combined_undirected}) and the six directed networks (Fig.~\ref{fig:combined_directed}) with eight neurons, $N = 8$. Each network has the form of a one-dimensional chain; in most cases the chain is supplemented by a set of shortcuts.

\subsection{Calculation of ITMs for evolving systems}

The conditional entropy (\ref{it6}) is expressed through the solution of the master equation (\ref{SND2c}) as follows
\begin{equation}
H_c(V_0|V)=\sum_{\vec v,\vec v_0} P(\vec v,t;[\delta_{\vec v\vec v_0}])P_0(\vec v_0)\log\frac{P(\vec v,t;[P_0(v)])}{P(\vec v,t;[\delta_{\vec v\vec v_0}])P_0(\vec v_0)}\ ,                                  \label{nsti4} 
\end{equation}
where $P_0(\vec v)$ is an arbitrary initial distribution, and $\delta_{\vec v\vec v_0}$ is such initial distribution that the configuration $\vec v_0$ is realized with the unitary probability:
\begin{equation}
 P(\vec v, t=0)\equiv P_0(\vec v)=\delta_{\vec v\vec v_0}\equiv\left\lbrace\begin{array}{lll} 1 & \mbox{if} & \vec v=\vec v_0\\[3mm] 0 & \mbox{if} & \vec v\neq\vec v_0 \end{array}\right. \label{nsti1}
\end{equation}

As can be seen from the expression (\ref{nsti4}), for the calculation of the conditional entropy and hence the AII it is necessary to find a solution of the system of equations (\ref{SND2c}) for $2^8 = 256$ initial conditions $\vec v_0$ plus solution for the actual initial distribution $P_0(\vec v)$. As concerns the latter it is worth mentioning that
\begin{equation}
 \sum_{\vec v_0}P(\vec v,t;[\delta_{\vec v\vec v_0}])P_0(\vec v_0)=P(\vec v,t;[P_0(v)])\ ,\label{nsti2}
\end{equation}
so knowing the solutions for $\delta_{\vec v\vec v_0}$ with all $\vec v_0$, the solution for an arbitrary initial distribution can be calculated by using (\ref{nsti2}). As the initial condition for the master equation we will use the uniform distribution: $P_0(\vec v)=P^N_{unif}=1/2^N\ \forall\, \vec{v}$.

\subsection{Calculation of the CMs for the neural networks under consideration}
Averages of the variables of neuronal activity can be calculated using the moment equation hierarchy using the methods of the stochastic neurodynamics \cite{OC}. If we needed only these averages, this way would be easier from a computational point of view. However, since a computation of the ITMs requires knowledge of the full distribution function of network configurations, it is certainly easier to calculate averages based on the known distribution. Given that the neural variables take the values $v_i=\{0,1\}$, the expressions for the moments reads
\begin{equation}
 \lgl v_{i_1}v_{i_2}\cdots v_{i_k}\rgr=\sum_{\stackrel{\scriptstyle{v_1,\dots, v_{i_1-1},v_{i_1+1}\dots,}}{v_{i_2-1},v_{i_2+1}\dots, v_{i_k-1},v_{i_k+1}\dots, v_{N}}}P_{v_1,\dots, v_{i_1-1},1,v_{i_1+1}\dots, v_{i_2-1},1,v_{i_2+1}\dots, v_{i_k-1},1,v_{i_k+1}\dots, v_{N}}\ .  \label{ksii1}
\end{equation}
In other words, for the averaging variable one has to take the probability of unit value for this variable and sum up over values of other neurons. In particular, for the uniform distribution one easily obtains
\begin{equation}
 \lgl v_{i_1}v_{i_2}\cdots v_{i_k}\rgr_{unif}=\frac{2^{N-k}}{2^N}=\frac{1}{2^k} \ .  \label{ksii1a}
\end{equation}

\subsection{The dependence of the conditional entropy on the average activity and the second moment of the neuronal activity}
It is well known that the value of integrated information determines ability of a systems to  the differentiation (discrimination/``recognition'' of the initial configuration) and the integration (higher capacity for differentiation of the entire system compared with the ensemble of its independent subsystems; in other words, global cohesion of the system). Conditional entropy is the basis for the calculation of the AII (see (\ref{bs7})) and determines the ability of the system to the differentiation of initial states. Roughly speaking, the larger the conditional entropy at a given time, the more neural network has ``forgotten'' by this time the initial configuration from which the evolution began. Therefore it is very useful to establish an interrelation of this ITM with correlation properties of networks.

\begin{figure}[t]
\begin{center}
\includegraphics[scale=0.7]{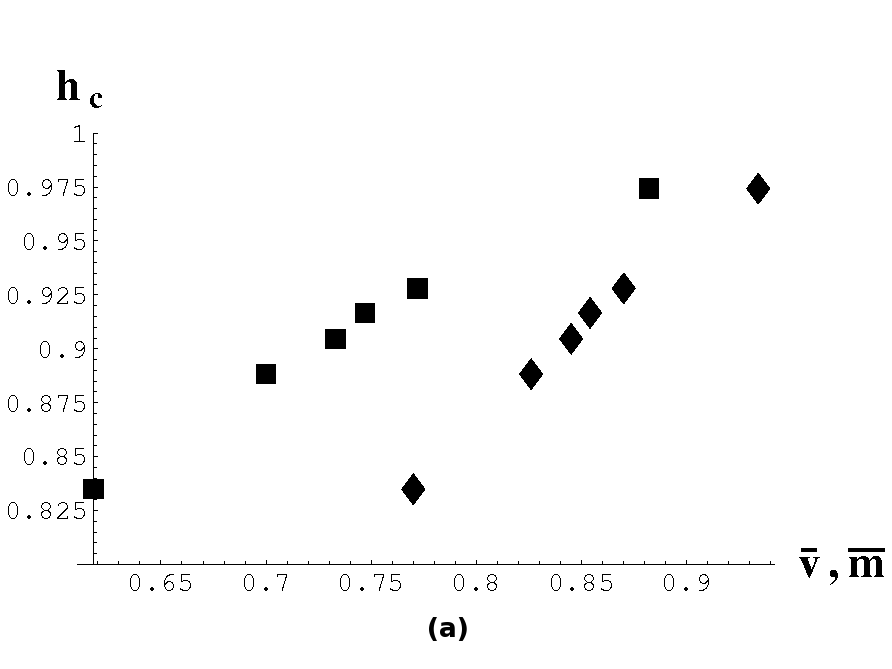}
\includegraphics[scale=0.7]{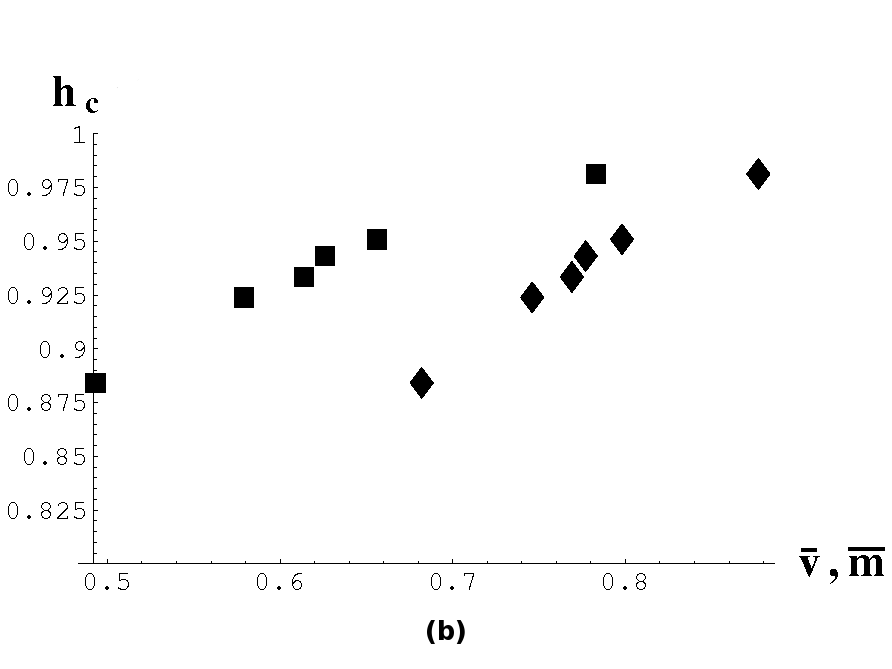}
\includegraphics[scale=0.7]{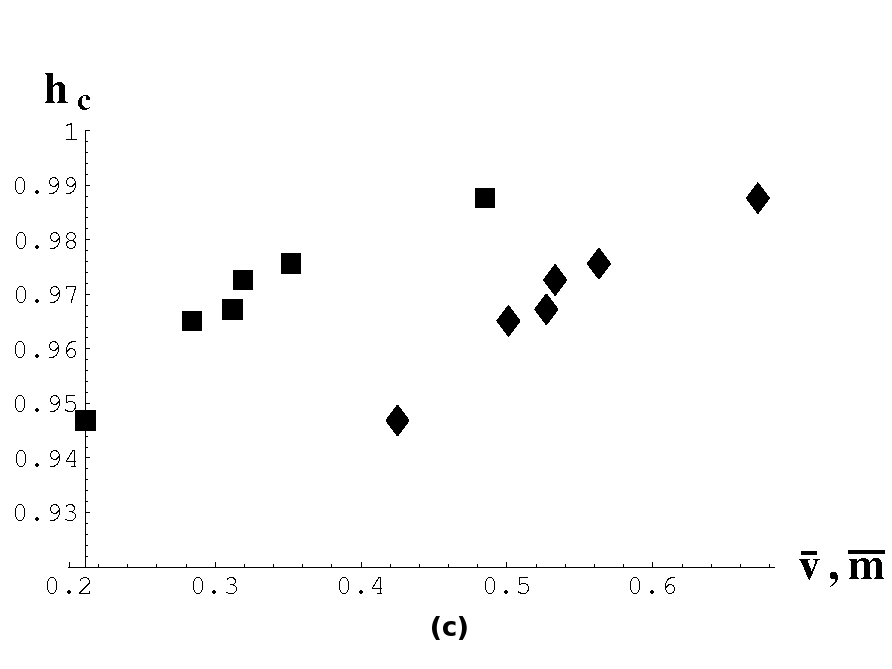}
\includegraphics[scale=0.7]{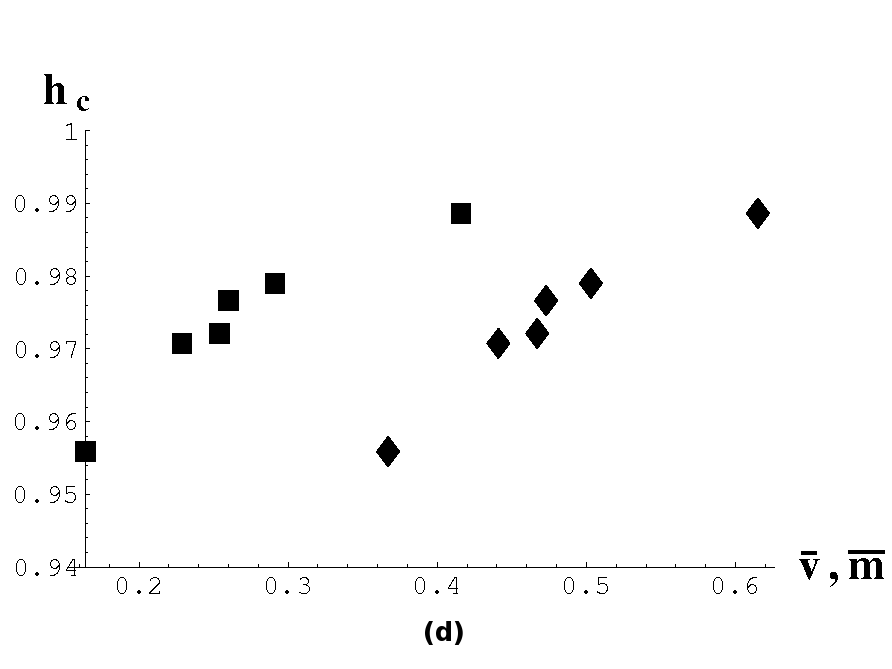}
\end{center}
\caption{The dependence of the normalized conditional entropy $h_c$ on values of the averaged activity $\bar{v}$  (triangles) and the second moment $\bar{m}$ (boxes) for the undirected networks depicted in the Fig.~\ref{fig:combined_undirected}: (a) $\lambda=0.1$; (b) $\lambda=0.3$; (c) $\lambda=1.0$; (d) $\lambda=1.2$ }
\label{fig:vm-H_undir_combined}
\end{figure}

Obviously, the value of the conditional entropy for binary networks is limited to a maximum value of $N$. Therefore it is convenient to present the results on the interrelation for the normalized conditional entropy
\begin{equation}
h_c(V_0|V_t)\deff H_c(V_0|V_t)/N\ .                  \label{nce1}
\end{equation}
The dependence of $h_c(V_0|V_{t=1})$ on the averaged activity $\bar{v}(t=1)$ (see (\ref{skh1})) and on the average second moment  $\bar{m}(t=1)$ (see (\ref{skh1a})) is presented in Fig.~\ref{fig:vm-H_undir_combined}. The figures show the values at the time $t=1$. Notice that in accordance with (\ref{ksii1a}), $\bar{v}(t=0)=1/2;\ \bar{m}(t=0)=1/4$. 

\begin{figure}
\begin{center}
\includegraphics[scale=0.7]{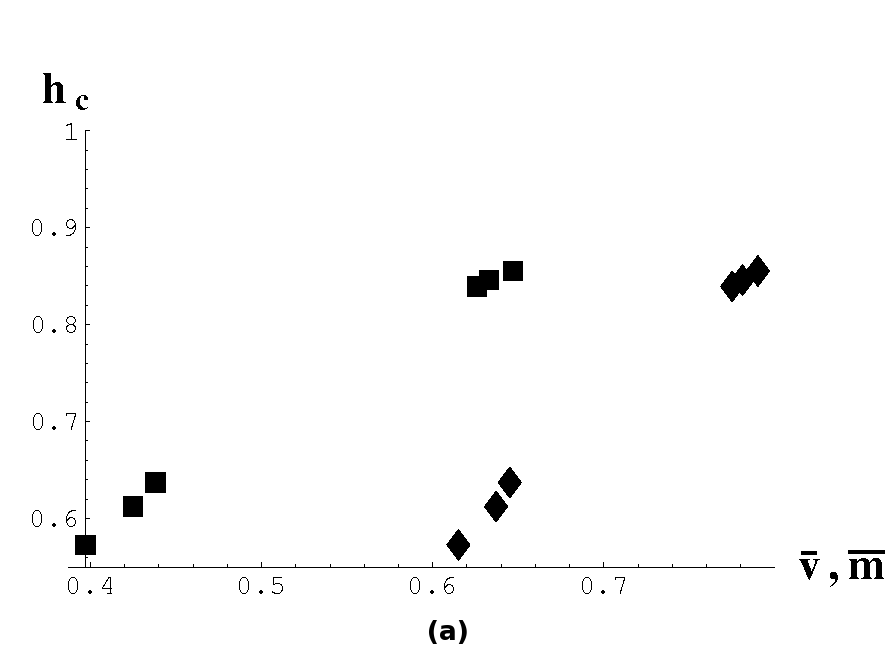}
\includegraphics[scale=0.7]{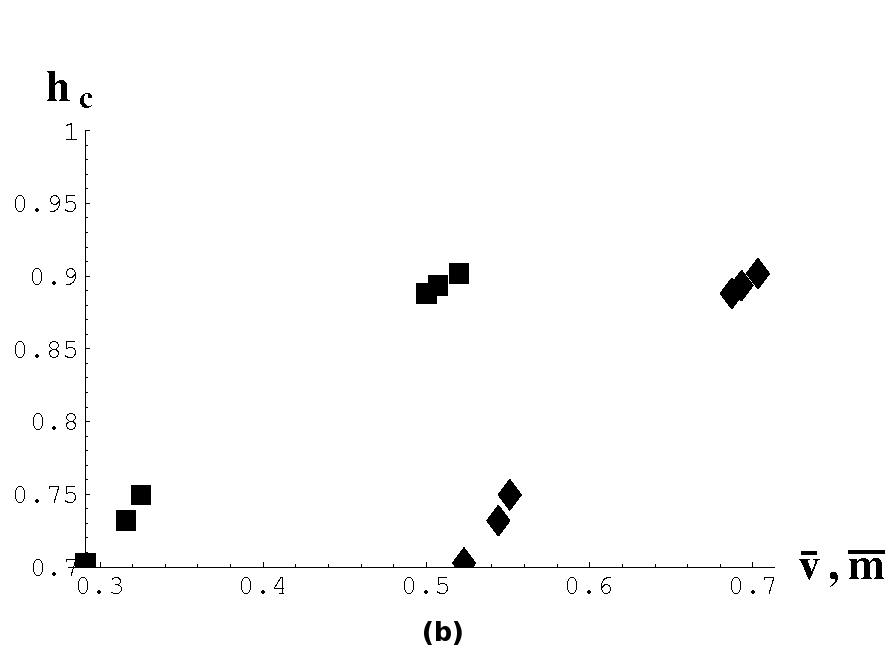}
\end{center}
\caption{The dependence of the normalized conditional entropy $h_c$ on values of the averaged activity $\bar{v}$  (triangles) and the second moment $\bar{m}$ (boxes) for the directed networks depicted in the Fig.~\ref{fig:combined_directed}: (a) $\lambda=0.1$; (b) $\lambda=0.3$}
\label{fig:vm-H_dir_combined}
\end{figure}

It is seen that if to fix the value of $\lambda$ and compare $h_c(V_0|V_{t=1})$ for different networks of the same type (directed or undirected), the conditional entropy increases monotonically with increasing values of $\bar{v}(t=1)$ and $\bar{m}(t=1)$ for a given network.

\begin{figure}
\begin{center}
\includegraphics[scale=1.0]{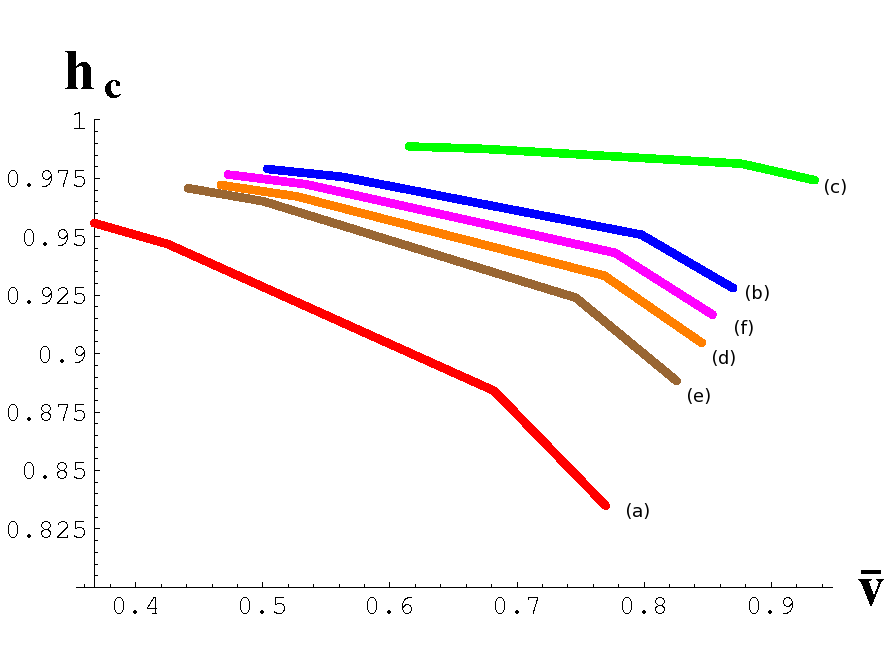}
\end{center}
\caption{The dependence of the normalized conditional entropy $h_c$ on values of the averaged activity $\bar{v}$ for the undirected networks and different values of $\lambda$; the letters next to the curves correspond to the designation of the networks in Fig.~\ref{fig:combined_undirected} (color on-line: (a) -- red; (b) -- blue; (c) -- green; (d) -- orange; (e) -- brown; (f) -- magenta)}
\label{fig:vH_var_ld_undir}
\end{figure}

On the contrary, from the plots in Fig.~\ref{fig:vH_var_ld_undir} it can be seen that for a given network and the different values of $\lambda$, the average values $\bar{v}(t=1)$, $\bar{m}(t=1)$ on the one side and $h_c(V_0|V_{t=1})$ on the other side are in an inverse relationship. Further discussion of these properties see in Sec.~\ref{sec:Zak}.

The dependence of the conditional entropy of the other correlation metrics are not so obvious or absent. For example, the average  Pearson coefficient $\bar{r}$ is almost the same for all undirected graphs in Fig.~\ref{fig:combined_undirected}, that is, its value is practically independent of the structure of the networks. This fact was mentioned in \cite{PSCR} (but without evidences, arguments or results of numerical calculations). The dependence of the conditional entropy on the average covariance are not stable and varies with values of $\lambda$; therefore, the latter CM is also inappropriate for establishing interrelations with ITMs.

\subsection{Integrated Information, Average Activity and Inter-Cluster Correlation Coefficient \label{sec:IIAK}}
\begin{figure}[t]
\begin{center}
\includegraphics[scale=0.7]{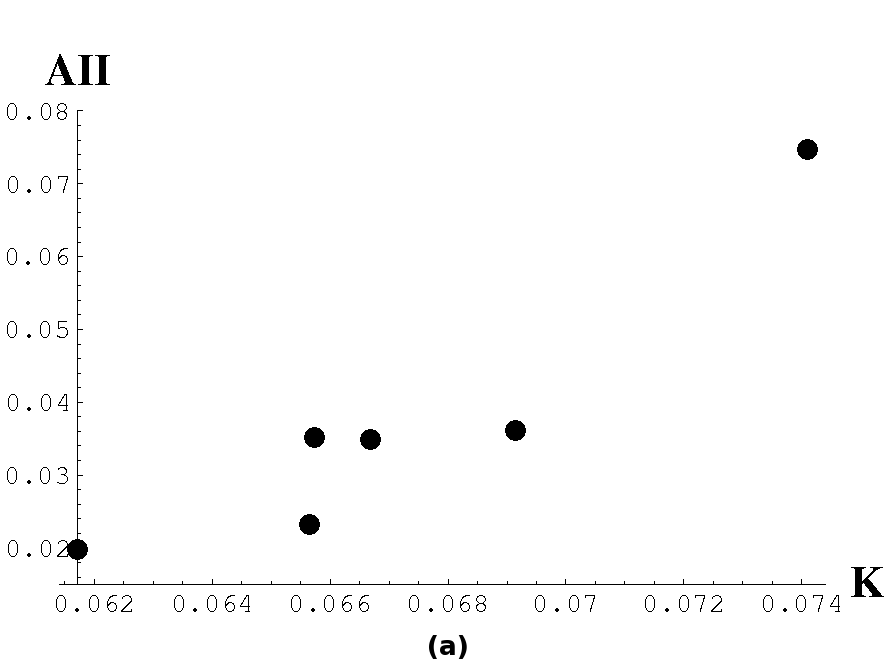}
\includegraphics[scale=0.7]{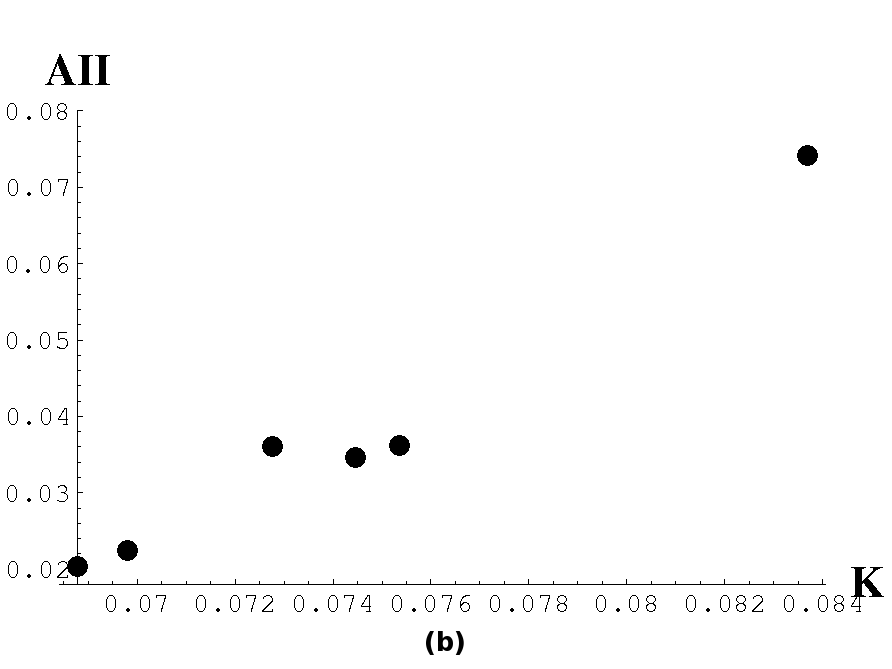}
\includegraphics[scale=0.7]{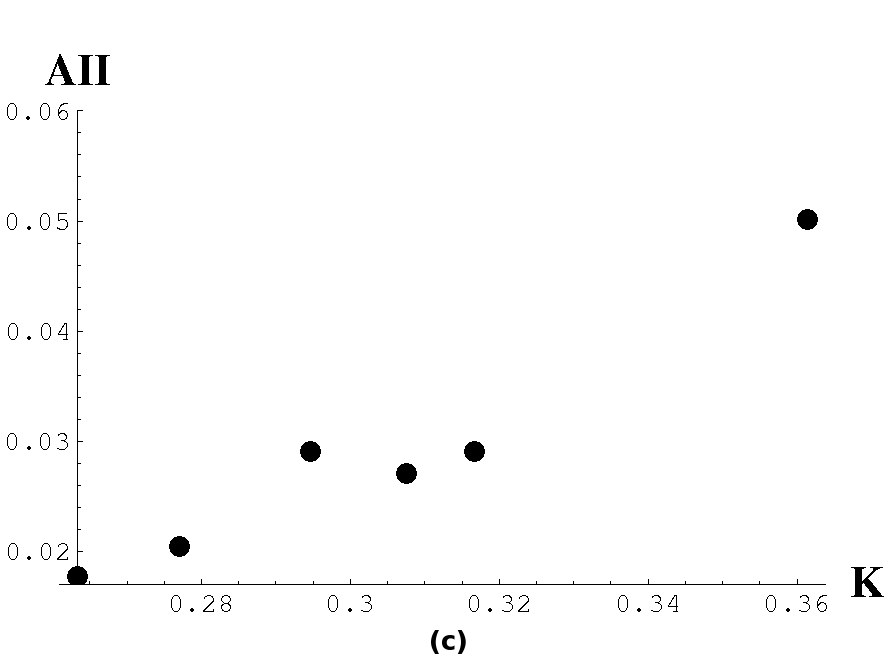}
\includegraphics[scale=0.7]{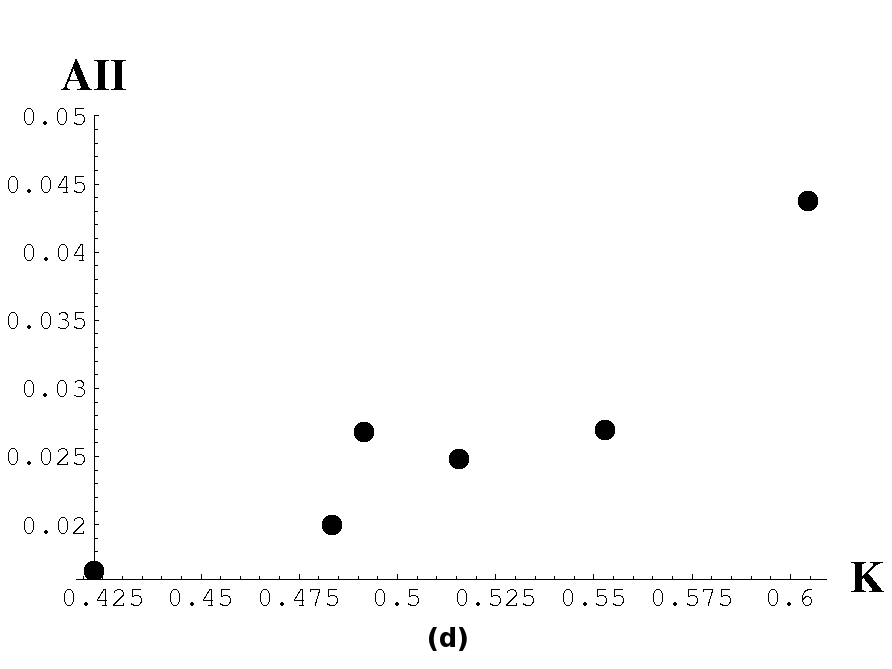}
\end{center}
\caption{The dependence of the average integrated information (AII) $\tilde{\Phi}$ on the correlation capacity for information integration (CCII) $K_\delta$ for the undirected networks in Fig.~\ref{fig:combined_undirected}: (a)~$\lambda=0.1,\ \delta=15$; (b)~$\lambda=0.3,\ \delta=15$; (c)~$\lambda=1.0,\ \delta=4$; (d)~$\lambda=1.2,\ \delta=2$}
\label{fig:AII-K_undirect_delta15-15-4-2}
\end{figure}

\begin{figure}
\begin{center}
\includegraphics[scale=0.7]{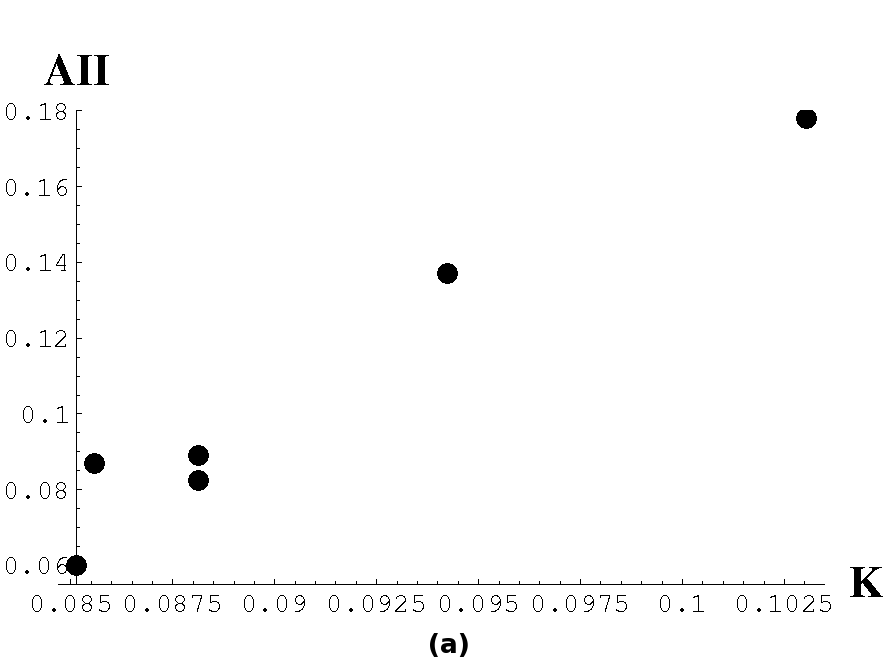}
\includegraphics[scale=0.7]{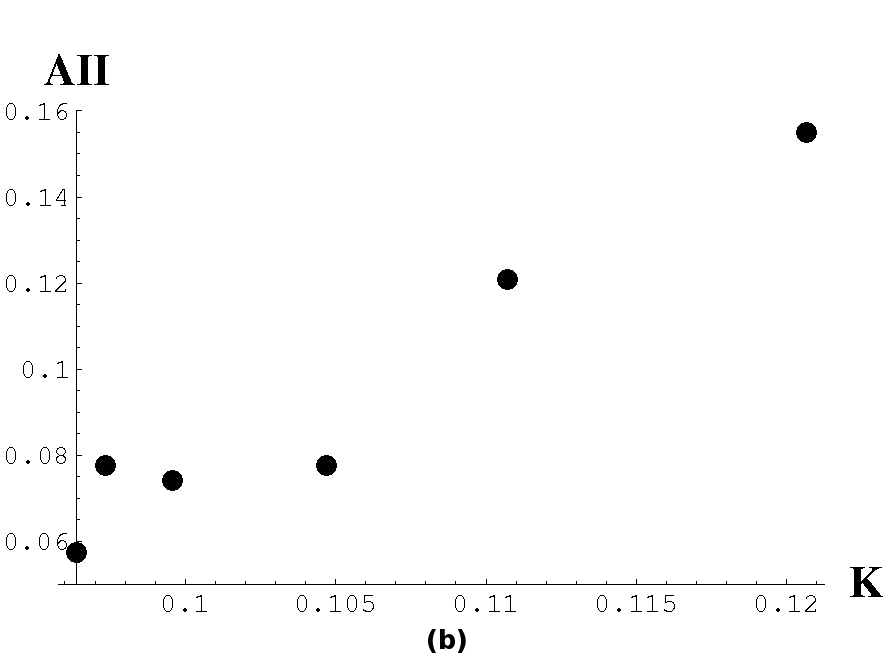}
\end{center}
\caption{The dependence of the average integrated information (AII) $\tilde{\Phi}$ on the correlation capacity for information integration (CCII) $K_\delta$ for the directed networks in Fig.~\ref{fig:combined_directed}: (a)~$\lambda=0.1,\ \delta=12$; (b)~$\lambda=0.3,\ \delta=12$}
\label{fig:AII-K_direct_delta12}
\end{figure}

The ability to differentiate the initial configurations at time $t = 1$ is the larger, the smaller the value of the conditional entropy $H_c(V_0|V_{t=1})$. In turn, $H_c(V_0|V_{t=1})$, as shown in the preceding section, is proportional to the values of $\bar{v}$ and $\bar{m}$, the entropy $H_c(V_0|V_{t=1})$ being more sensitive to variations of the average activity than to $\bar{m}$. Therefore, we can assume that the value of the integrated information is inversely proportional to the $\bar{v} $. The integration of information across the system should depend on the interconnection between its subsystems. As a measure of the interconnection of individual subsystems we choose the inter-cluster correlation coefficient (\ref{WC1b}). Obviously, if the network consists of two (or more) disconnected parts $S$ and $\hat{S}$, then ICC is zero (all $c_{ij},\ i\in S,\ j\in\hat{S}$ are equal to zero). AII for such neural network is also, obviously, must be zero (since there is a decomposition into subsystems (disconnected components) that differentiate the initial state not worse than the whole network). We can therefore assume that the smaller the minimal inter-cluster correlation coefficient $k\deff \min_{S \subset V} k_{S\hat{S}}$, the smaller should be the AII. These arguments prompt entering the following quantity, which we call the correlation capacity for information integration (CCII)
\begin{equation}
 K\deff \frac{1}{\bar{v}}\cdot\frac{1}{\delta +1/k}\ ,   \label{itcq1}
\end{equation}
where $\delta$ is a phenomenological parameter. The relationship between CCII and AII for the neural networks depicted in Fig.~\ref{fig:combined_undirected} and \ref{fig:combined_directed}, is shown in Fig.~\ref{fig:AII-K_undirect_delta15-15-4-2} and \ref{fig:AII-K_direct_delta12}, respectively. It is seen that the magnitude of the AII strongly correlates with CCII and the dependence for the selected values of the parameter $\delta$ is close to a linear one.

\section{Discussion and conclusion\label{sec:Zak}}
In this paper we studied the relationship between such a complex information-theoretic metric as the integrated information (more precisely, the average integrated information), and simpler correlation characteristic properties of neural networks, namely, average activity, the correlation of activities of individual neurons and inter-cluster correlations. The advantage of the latter metrics is that they are expressed only in terms of the moments of neuron activities of finite order so that it is much easier to develop approximate methods for their calculations in the case of large realistic networks. As mentioned in the Introduction, an establishment of exact interdependence by analytical methods is a very complicated problem. Therefore in this study we tried to establish a quantitative relationship of these metrics for a certain set of neural networks of small size, namely, with eight neurons $N = 8$. Even for such a relatively small network computing such information-theoretic metrics as the conditional entropy and average integrated information (AII) requires the solution of a large number of equations which however can be done using computer algebra systems. As a result, the dependence of conditional entropy and AII on neuron correlations has been established for such networks.

As a model of neural networks the stochastic neurodynamics \cite {Cow}, \cite {BC} was chosen. We believe that since real neural networks are subject to internal and external random influences (see., e.g., \cite{SK}, \cite{PTW}, \cite{FSW}), stochastic methods are required for their description. The variant of the stochastic neurodynamics chosen in this work  is quite simple and suitable for numerical analysis and, at the same time, retains some important qualitative features of real neural networks.

The main results obtained in this work are:
\begin{itemize}
\item establishing a direct dependence (Fig.~\ref{fig:vm-H_undir_combined} and Fig.~\ref{fig:vm-H_dir_combined}) between the value of the average activity $\bar{v} $ and the conditional entropy $H_c(V_0|V_{t=1})$ for the same values of the ratio $\lambda$ of deactivation and activation constants and for various neural networks of a given type (undirected or directed) ;
\begin{itemize}
\item a similar direct dependence exists also for conditional entropy and the average second moment of activities $\bar{m}$, but  dependence of the $H_c$ on the average activity $\bar{v}$ is stronger;
\item these dependencies are approximately linear;  
\end{itemize}
\item for a fixed network structure the conditional entropy is inversely dependent (Fig.~\ref{fig:vH_var_ld_undir}) on the magnitude of the average activity (which, in turn, is greater for smaller values of the ratio $\lambda$ of deactivation and activation constants);
\item out of $\bar{v}$ and the minimal inter-cluster correlation coefficient $k$ it is possible to construct the quantity $K_\delta$, which we called a correlation capacity for information integration, such that with a suitable choice of the free phenomenological parameter $\delta$ the average integrated information approximately linearly depends on $K_\delta$  (Fig.~\ref{fig:AII-K_undirect_delta15-15-4-2} and Fig.~\ref{fig:AII-K_direct_delta12}).
\end{itemize}

\begin{figure}
\begin{center}
\includegraphics[scale=0.9]{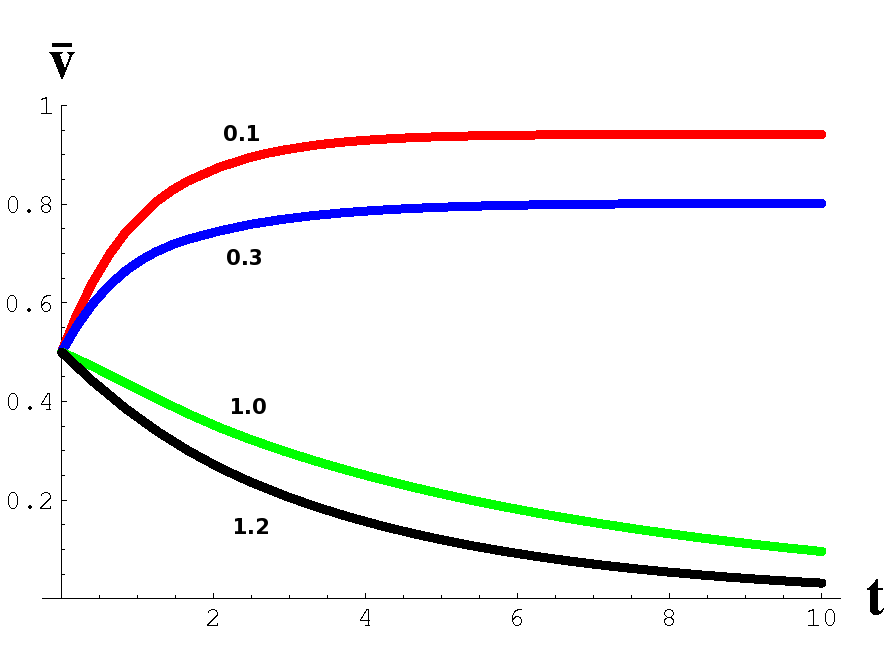}
\end{center}
\caption{The evolution of the average activity of the neural network (a) in Fig.~\ref{fig:combined_undirected} for various values of the parameter $\lambda$ (color on-line): $\lambda=0.1$  (red), $\lambda=0.3$ (blue), $\lambda=1.0$ (green), $\lambda=1.2$ (black curve) (the numbers next to the curves correspond to the values of the parameter $\lambda$ )}
\label{fig:vEvolution}
\end{figure}

It is worth mentioning that there exist hypotheses in the literature that biological neural networks most effectively process information in metastable states, i.e. near phase transition points (see., eg, \cite{FBFC}, \cite{Wer}). Of course, in the case of finite-size networks it is impossible to observe a genuine phase transition. But such a transition was detected in a similar model for a large (actually infinite) closed chain of neurons \cite{OC}: steady-state network for small $\lambda$ is fully active, while for large $\lambda$ it is completely quiescent (inactive). In our notation, the  change of the steady states occurs at $\lambda = 1$ for the infinite chain. However, even in small networks one can see a trace of such a ``phase transition'', namely, the change of sign of the time derivative at $t = 0$ of the average activity at some value of $\lambda$. In other words, for large $\lambda$ the network activity from the very beginning of the evolution monotonically falls down while for small $\lambda$ the activity  at first  increases and only then begins to fall (for infinitely large network totally active configuration may become a stationary one \cite{OC}). Such an alteration in the system behavior can be considered as a trace of the phase transition in the case of the small systems. The range of $\lambda$ where the average activity at the initial time is almost constant can be considered as an analogue of the phase transition point. Examples of the average activity evolution for a number of values of $\lambda$ are shown in Fig.~\ref{fig:vEvolution}. 

\begin{figure}
\begin{center}
\includegraphics[scale=0.9]{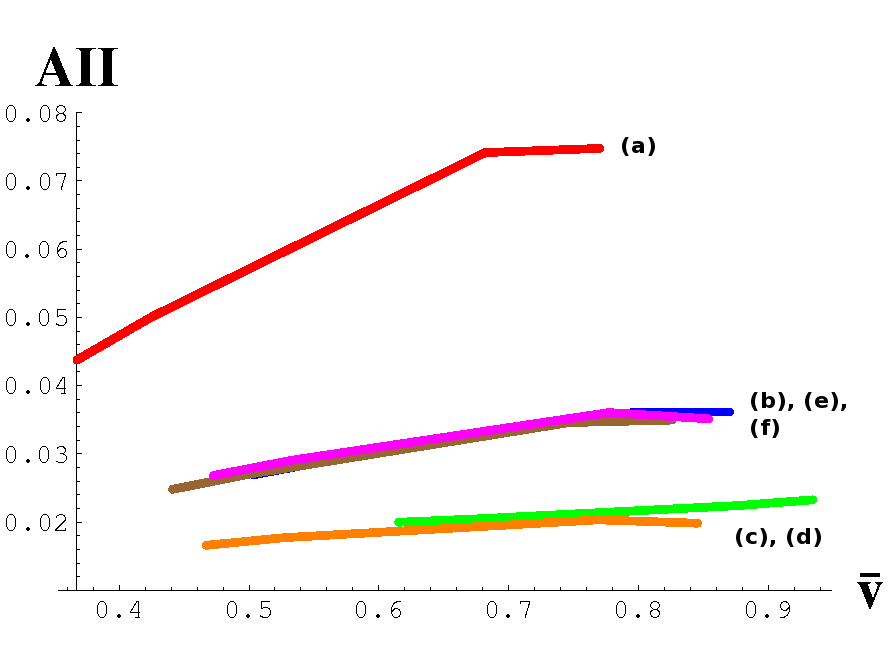}
\end{center}
\caption{The dependence of the average integrated information on the average activity $\bar{v}$ for the undirected networks and various values of  $\lambda$; the letters next to the curves correspond to the  designation of the networks in Fig.~\ref{fig:combined_undirected} (color on-line: (a) -- red; (b) -- blue; (c) -- green; (d) -- orange; (e) -- brown; (f) -- magenta)}
\label{fig:vAII_sameNN.png}
\end{figure}

We could assume that the system has the highest AII near the area of ``phase transition'' where the network is not very active and not very relaxed; cf. also the arguments in \cite{BT2008}. However, our numerical results show that this assumption is not justified: the AII monotonically grows with growing the average activity, see. Fig.~\ref{fig:vAII_sameNN.png} (recall that the initial value of $\bar{v}=0.5$).

Notice also that similarly to the average activity (Fig.~\ref{fig:vEvolution}), by means of solving the equation (\ref{SND2c}) one can compute values of any other quantities not only at $t = 1$ but also for an arbitrary value of $t$, that is to determine their total evolution. In particular, it is rather curious that the MIBs, used for calculation of the integrated information, vary in the evolution process. As far as we know, there is no mentioning of this fact in the literature.

We hope that in the future the results obtained in this work can be generalized to certain classes of large size neural networks. In particular, in subsequent studies a specific class of two-layer networks whose basic layer consists of regular short-range links (such as in regular lattices or fractals) supplemented by the second layer of long range shortcuts will be  	investigated. Such networks can be considered as a prototype for cognitive neural networks because a similar hierarchical structure was previously observed in experimental studies of structural and functional connections in human brain. In addition, the results obtained in this paper for the small size networks may be of independent interest, for example when using the information-theoretic metrics for developing neuromorphic robots.

\end{document}